\documentclass[11pt]{article}
\usepackage{latexsym,amssymb,amsmath}
\usepackage[dvips]{graphics,graphicx,epsfig,color}
\begin{document}
\thispagestyle{empty}
\begin{center}
 {\bf\large AMPLIFICATION AND INCREASED DURATION OF EARTHQUAKE MOTION ON UNEVEN STRESS-FREE GROUND}
\end{center}
\begin{center}
{Armand WIRGIN$^{1}$,  Jean-Philippe GROBY$^{2}$\\ {\it $^{1}$
Laboratoire de M\'ecanique et d'Acoustique, UPR 7051 du CNRS,
Marseille, France.\\$^{2}$ Laboratorium v. Akoestiek en Thermische
Fysica,  Katholieke Universiteit, Leuven, Belgium.}}
\end{center}
~~ \\ ~~~
%
%
{\bf ABSTRACT-} When a flat stress-free surface (i.e., the ground
in seismological applications) separating air from a isotropic,
homogeneous or horizontally-layered, solid substratum is solicited
by a SH plane body wave incident in the substratum, the response
in the substratum is a single specularly-reflected body wave. When
the stress-free condition, equivalent to vanishing surface
impedance, is relaxed by the introduction of a {spatially-
constant, non- vanishing surface impedance},  the response in the
substratum is again a single reflected body wave whose amplitude
is less than the one in the situation of a stress-free ground.
When the stress-free condition is relaxed by the introduction of a
 a {spatially-modulated surface impedance},
 which simulates the action of an
uneven (i.e., not entirely-flat) ground , the frequency-domain
response takes the form of a  spectrum of {plane body waves} and
{surface waves} and {resonances} are produced at the frequencies
of which one or several surface wave amplitudes can become large.
It is shown, that at resonance, the amplitude of one, or of
several, components of the motion on the surface can be amplified
with respect to the situation in which the surface impedance is
either constant or vanishes. Also,  when the solicitation is
pulse-like, the integrated time history of the square of surface
displacement and of the square of velocity can be larger, and the
duration of the signal can be considerably longer, for a
spatially-modulated impedance surface than for a constant, or
vanishing, impedance surface.
\newline
\section{Introduction}\label{genin}
An important question in seismology, civil engineering, urban
planning, and natural disaster risk assessment is: to what extent
does surface topography of different length and height scales
(ranging from those of mountains and hills to city blocks and
buildings) modify the seismic response (in terms of cumulative
motion intensity and duration) on the ground?

There exists  {experimental evidence} \nocite{sior93}(Singh and
Ordaz, 1993;  \nocite{dawe73}Davis and West, 1973;
\nocite{grbo79}Griffiths and Bollinger, 1979) that this
modification is real and can attain considerable proportions.

Some  {theoretical studies} (\nocite{wi89}Wirgin, 1989;
 \nocite{wi90b}Wirgin 1990;
\nocite{wiko93}Wirgin and Kouoh-Bille, 1993; \nocite{gr05}Groby,
2005) seem to indicate that such effects are indeed possible, but
various  {numerical studies} ( \nocite{bo73} Bouchon, 1973;
\nocite{ba82}Bard, 1982; \nocite{sase87}Sanchez-Sesma, 1987,
\nocite{geba88}Geli et al., 1988; \nocite{wiba96}Wirgin and Bard,
1996; \nocite{gu00}Gu\'eguen, 2000; \nocite{clau01}Clouteau and
Aubry, 2001; \nocite{guba02}Gu\'eguen et al., 2002;
\nocite{segu03}Semblat et al., 2003; \nocite{tswi03}Tsogka and
Wirgin, 2003;  \nocite{boro04}Boutin and Roussillon, 2004;
\nocite{kh04}Kham, 2004; \nocite{grts05}Groby and Tsogka, 2005)
yield  {conflicting results} in that some of these point to
amplification, while others to very weak effects, or even to
de-amplification. Contradictory results are also obtained
regarding the duration of the earthquakes.
\subsection{Sites}
\begin{figure}[ptb]
\begin{center}
\includegraphics[scale=0.25] {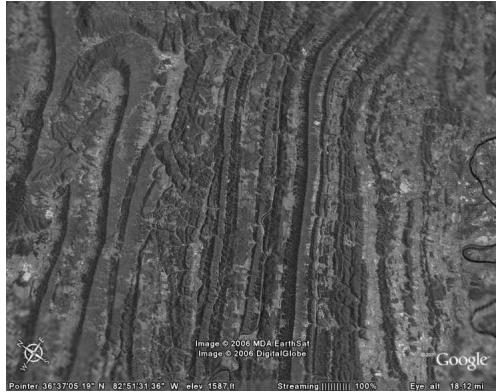}
\end{center}
\caption{Appalachian mountains.} \label{one}
\end{figure}

\begin{figure}[ptb]
\begin{center}
\includegraphics[scale=0.25] {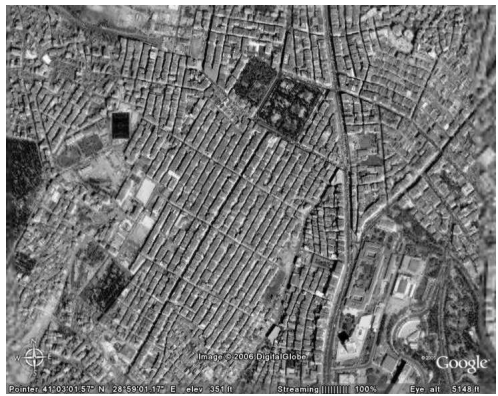}
\end{center}
\caption{Istanbul, Turkey.}
 \label{two}
\end{figure}

In  figs. \ref{one} and \ref{two} we give  examples of natural and
man-made sites respectively with quasi-periodic features that can
be studied by the methods of this investigation.
\subsection{Physical configurations}
\begin{figure}[ptb]
\begin{center}
\includegraphics[scale=0.45] {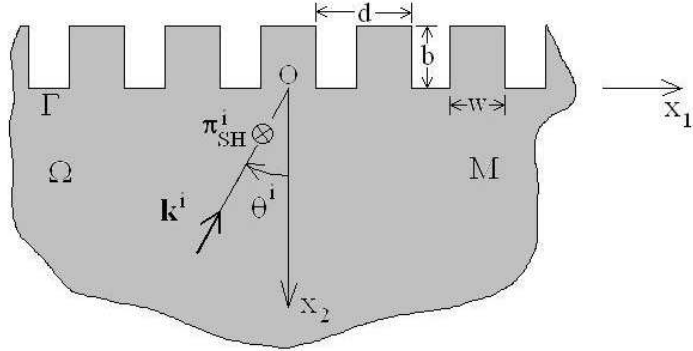}
\end{center}
\caption{Periodically- uneven ground.}
 \label{perunev}
\end{figure}
\begin{figure}[ptb]
\begin{center}
\includegraphics[scale=0.45] {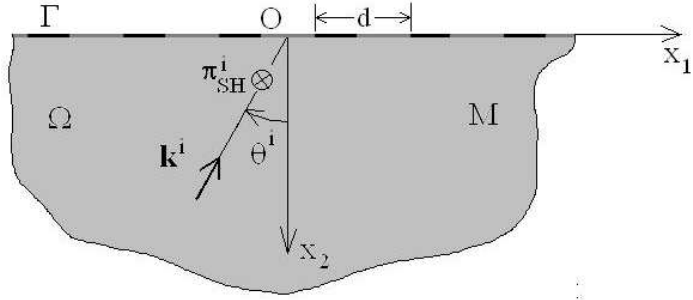}
\end{center}
\caption{Periodically- modulated impedance  flat ground.}
\label{periimp}
\end{figure}
 The (assumed) periodic uneveness of the ground (an example of which is given in fig. \ref{perunev}) is
accounted for by a suitably-chosen {spatially-periodic flat
surface impedance function} (see fig. \ref{periimp}).
\subsection{On the notion of impedance}
We shall employ the notion of (surface) {imped}{ance} employed in
the civil engineering \nocite{gu00}(Gu\'eguen, 2000),
\nocite{ro06}(Roussillon, 2006) community wherein the three
components of impedance are $M,~K,~C$ which designate mass,
stiffness and damping ( real constants) respectively.

Thus, the mechanical impedance is complex and designated by
$Z=R-iX$, where {$R=C$ is the "resistive" part}, and {$X=\omega
M-K/\omega$ the "reactive" part}.

The {reactance is "inductive" if $\omega^{2}> K/M$ (i.e., $X>0$)}
and is {"capacitive" if $\omega^{2} < K/M$ (i.e., $X<0$}.
\section{Governing equations}
%
\subsection{Mathematical translation of the boundary value
problem in the space-frequency domain}
\begin{equation}\label{onehunthirteen}
  \mu [u_{,11}(\mathbf{x},\omega)+u_{,22}(\mathbf{x},\omega)] +
    \rho\omega^{2} u(\mathbf{x},\omega)=0~;
    \forall\mathbf{x}\in\Omega
 ~,
\end{equation}
\begin{equation}\label{onehunthirteenab}
 i\omega Z(x_{1},\omega)u(\mathbf{x},\omega)+\mu u_{,2}(\mathbf{x},\omega)=0~;
 ~\forall\mathbf{x}\in\Gamma~,
\end{equation}
\begin{equation}\label{onehunfourteenc}
u^{d}(\mathbf{x},\omega):=
u^{0}(\mathbf{x},\omega)-u^{i}(\mathbf{x},\omega)\sim
\text{outgoing
waves}~;~\|\mathbf{x}\|\rightarrow\infty~~,~~\mathbf{x}\in\Omega~,
\end{equation}
\begin{equation}\label{onehunfourteene}
u^{i}(\mathbf{x},\omega)=A^{i}(\omega)\exp[i(k_{1}^{i}x_{1}-k_{2}^{i}x_{2})]~;
~\forall\mathbf{x}\in\Omega~~, \\~~k_{1}^{i}=
k\sin\theta^{i}~,~~k^{i}_{2}=k\cos\theta^{i}~,~~k=\frac{\omega}{c}~,
\end{equation}
wherein: $u$ is the total displacement field, $u^{d}$ the
(unknown) diffracted field, $u^{i}$ the (known) incident field ,
$\theta^{i}$ the angle of incidence with respect to the $x_{2}$
axis, $A^{i}(\omega)$ the incident pulse spectrum.
\begin{equation}\label{onehunfourteenff}
Z(x_{1},\omega):=R(x_{1},\omega)-iX(x_{1},\omega)~,
\end{equation}
When the impedance vanishes for all $x_{1}$, the boundary
condition becomes that of a flat, stress-free surface.

Otherwise, the impedance boundary condition is supposed to
simulate the presence of a topo-graphically- uneven stress-free
surface.

The periodic nature of $Z$ is expressed by:
\begin{equation}\label{twohunfiftyfive}
Z(x_{1}+d,\omega)=Z(x_{1},\omega)~~;~~\forall x_{1}\in
\mathbb{R}~,
\end{equation}
\begin{equation}\label{twohunfiftyseven}
Z_{l}=\int_{-\frac{d}{2}}^{\frac{d}{2}}Z(x_{1})\exp\left(
-i\frac{2l\pi}{d}x_{1}\right) \frac{dx_{1}}{d}~~;~~\forall l\in
\mathbb{Z}~.
\end{equation}
It is also assumed that  $R(x_{1},\omega)> 0$, i.e., the impedance
is  passive and dissipative.
\subsection{Diffracted field representation}
\begin{equation}\label{twohunseventytwo}
u^{d}(x_{1},x_{2})=\sum_{n=-\infty}^{\infty}A_{n}
\exp[i(k_{1n}x_{1}+k_{2n}x_{2})]~;~\forall (x_{1},x_{2})\in
\Omega~,
\end{equation}
\begin{equation}\label{twohunseventyfour}
k_{1n}=k_{1}^{i}+\frac{2n\pi}{d}~,~k_{2n}=\sqrt{k^{2}-k_{1n}^{2}}~,~\Re
k_{2n}\geq 0~;~\Im k_{2n}\geq 0~,~\omega\geq 0~.
\end{equation}
The {diffracted field is  a discrete sum of plane waves}:  those
for which $k_{2n}$ is real are { propagative} (or homogeneous) and
those for which $k_{2n}$ is imaginary are {evanescent} or
(inhomogeneous).
\subsection{Result of the introduction of the field
representation into the boundary condition}
\begin{equation}\label{twohuneightyfour}
\sum_{n=-\infty}^{\infty} \left[
Z_{m-n}+\gamma_{n}\delta_{nm}\right]
B_{n}=-2Z_{m}A^{i}~~;~~\forall m\in \mathbb{Z}~,
\end{equation}
wherein
\begin{equation}\label{twohuneightythree}
B_{n}=A_{n}- A^{i}\delta_{n0}~~,~~\gamma_{n}:=\frac{\mu
k_{2n}}{\omega}~~;~~\forall n\in \mathbb{Z}~.
\end{equation}
The linear system can be written as the infinite-order {matrix
equation}
\begin{equation}\label{twohuneightytwob}
\mathbf{E}\mathbf{f}=\mathbf{g}~.
\end{equation}
\section{On the possibility of anomalous fields in the general
case of a periodic, passive, spatially non-constant, surface
impedance}
A {mode} of the configuration is obtained by turning off the
solicitation in the matrix equation, i.e.,
$\mathbf{g}=\mathbf{0}$, wherein $\mathbf{0}$ is the null vector.

We are thus faced with the equation
$\mathbf{E}\mathbf{f}=\mathbf{0}$, whose solution is trivial
(i.e., $\mathbf{f}=\mathbf{0}$), unless
\begin{equation}\label{threehuneight}
\text{det}(\mathbf{E})=0~.
\end{equation}
An "eigenvalue" is a value of $k_{1n}$ for which the determinant
vanishes at a given frequency. Another way of putting things is to
fix $k_{1n}$ and look for the frequencies ("natural frequencies")
that lead to a vanishing determinant.

When the configuration is such that one of the  $k_{1n}$ is an
eigenvalue, and the frequency is a natural frequency, then the
system is said to be in a state of  {resonance}.

When this happens, the determinant of $\mathbf{E}$ is either small
or nil, which means that the inverse of $\mathbf{E}$ is either
large or infinite and that consequently one or more entries in the
vector $\mathbf{f}$ of the scattered plane wave coefficients are
either large or infinite.

Consequently, we can expect (since the field is a discrete sum of
scattered plane waves) that {the field may become large at
resonance} (in the presence of not too much dissipation and/or
radiation damping).

From now on, we call det$(\mathbf{E})=0$ the {general dispersion
relation}. In the case of {spatially-constant} surface impedance
$Z_{0}=\zeta$, only the zeroth-order diffracted wave is generated,
so that  the dispersion relation is
\begin{equation}\label{threehunten}
 \zeta+\frac{\mu k_{2}^{i}}{\omega}=0~.
\end{equation}
and the latter has no solution for a passive impedance (i.e.,
$\Re\zeta\geq 0$) due to the fact that $k_{2}^{i}$ is real.

This means that {no resonance can be produced for a constant,
passive impedance surface}. When the surface impedance function
{{\it Z}  is not spatially-constant}, it is much more difficult to
obtain a meaningful expression of the dispersion relation.

Some insight may be gained by turning to an iteration method of
solution
\begin{equation}\label{ninehunthirtysixx}
B_{m,N}^{(l,N)}=\frac{-2A^{i}Z_{m}-\sum_{n=-N,\neq m}^{N}
Z_{m-n}B_{n}^{(l-1,N)}}{Z_{0}+\gamma_{m}}~~;
~~m=-N,..,0,..,N~;~l=1,2,....~.
\end{equation}
which suggests that $B_{m}$ can become large for
\begin{equation}\label{threehuneleven}
Z_{0}+\frac{\mu k_{2m}}{\omega}=0~.
\end{equation}
This cannot occur for $m=0$ for the previously-mentioned reason.
It can occur, if at all, only for $m\neq 0$. Recall that it was
assumed that $\Im \mu$=0, $\Im k=0$ and
\begin{equation}\label{threehuntwelve}
Z(x_{1})=R(x_{1})-iX(x_{1})~\Rightarrow~
Z_{m}(x_{1})=R_{m}(x_{1})-iX_{m}(x_{1})~.
\end{equation}
Consider the case of vanishing resistance, i.e., $R(x_{1})=0$ in
which the {approximate dispersion relation} is
\begin{equation}\label{threehunthirteen}
-iX_{0}+\frac{\mu} {\omega}k_{2m}=0~.
\end{equation}
Since $\omega\geq 0$, $\Re \mu>0$, $\Re k_{2m}>0$, and $\Re
k_{2m}>0$, the second term in the previous equation is either
positive real (for real $k_{2m}$) or positive imaginary (for
imaginary $k_{2m}$), so that the sum of the two terms can vanish
only if
\begin{equation}\label{threehunfourteen}
X_{0}>0~~~ \text{and}~~~\Re k_{2m}=0~.
\end{equation}
The first of these requirements means that the impedance must be
{\it inductive}  for it to be possible to obtain resonant
behavior. {Thus, a possible explanation of why several
researchers, who employed the impedance concept to account for
ground uneveness, have not been able to obtain anomalous fields is
that their impedance functions were such that $X_{0}\leq 0$ in the
frequency range of the incident pulse}. The second requirement,
i.e., $\Re k_{2m}=0$, means that {resonances can occur only for
the evanescent waves in the plane wave representation of the
scattered field}. Thus, we can expect the amplitude of the $m$-th
order evanescent wave to become infinite (for $R=0$) or large (for
$R>0$) at resonance, which is another way of saying that {a
surface wave (evanescent waves are of this sort) is strongly
excited at resonance (all the more so the smaller is $R$)}.

This picture is only partially true, because the approximate
dispersion relation may account only poorly for all the features
of the solutions of the general dispersion relation, which is the
case if the uneveness of the ground is large.

The fact that $B_{m}$ becomes large at resonance is a {necessary
condition} for the ground motion to be large at this resonance
frequency, but is {not a sufficient condition} due to the fact
that the diffracted field is composed not of one evanescent plane
wave, but of a sum of both propagative and evanescent plane waves,
and this sum can be of modest proportions even when one of its
components is large. {Such modest fields at resonance are due
mainly to radiation damping}.
\subsection{The case of spatially- sinusoidal
surface impedance}
For the computations, wee shall choose:
\begin{equation}\label{threehunfourteena}
Z(x_{1},\omega)=\zeta(\omega)\left[ 1+h\cos\left(
\frac{2\pi}{d}x_{1}\right) \right]~.
\end{equation}
When $h=0$, we re-encounter the case of a constant impedance on
flat ground, so that {$h$ is a measure of the uneveness of the
ground.}
\section{Computations}
We shall be interested in the following quantities indicative of
possible anomalous effects provoked by the uneveness of the
ground:
\begin{itemize}
\item the duration of significative seismic ground motion,
\item the peak value of the ground displacement,
\item the amplification factors (i.e., amplification if the factor $>1$, de-amplification if the factor $< 1$)
of the time integral (over $[0,\tau]$) of the squared displacement
$u$ (velocity v) for the modulated impedance surface
 relative to the time integral (over $[0,\tau]$) of squared displacement (velocity) for the
 $h=0$ impedance surface:
\begin{equation}\label{threehunfourteenb}
\chi(\zeta,h,\tau)=\frac{\int_{0}^{\tau}[u(0,0,t|\zeta,h)]^{2}dt}{\int_{0}^{\tau}[u(0,0,t|\zeta,0)]^{2}dt}
~~,~~\upsilon(\zeta,h,\tau)=\frac{\int_{0}^{\tau}[{\text
v}(0,0,t|\zeta,h)]^{2}dt}{\int_{0}^{\tau}[{\text
v}(0,0,t|\zeta,0)]^{2}dt}~,
\end{equation}
\item the amplification factors of the time integral (over
$[0,\tau]$)of squared displacement (velocity) for the modulated
impedance surface  relative to the time integral (over $[0,\tau]$)
of squared displacement (velocity) for the stress-free surface:
\begin{equation}\label{threehunfourteenc}
\eta(\zeta,h,\tau)=\frac{\int_{0}^{\tau}[u(0,0,t|\zeta,h)]^{2}dt}{\int_{0}^{\tau}[u(0,0,t|0,0)]^{2}dt}
~~,~~\nu(\zeta,h,\tau)=\frac{\int_{0}^{\tau}[{\text
v}(0,0,t|\zeta,h)]^{2}dt}{\int_{0}^{\tau}[{\text
v}(0,0,t|0,0)]^{2}dt}~.
\end{equation}
\end{itemize}
 The amplitude spectrum of
the incident plane wave is that of a Ricker pulse, i.e.,
$A^{i}(\omega)=-\frac{\omega^{2}}{4\alpha^{3}\sqrt{\pi}}
\exp(-\frac{\omega^{2}}{4\alpha^{2}}+i\omega\beta)$.

The other parameters are: $\alpha=1$, $\beta=4$, $s^{i}=0.3$,
$r=0.1$, $m=2$, $\kappa=1$, $s=2$.

\begin{figure}[ptb]
\begin{center}
\includegraphics[scale=0.6] {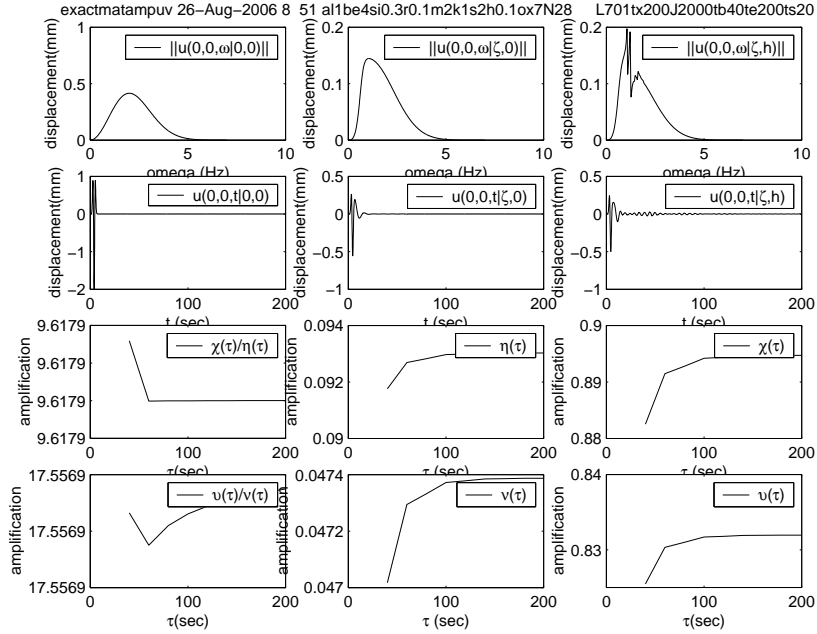}
\end{center}
\caption{$h=0.1$ . Small-scale anomalous effects.} \label{graf1}
\end{figure}

\begin{figure}[ptb]
\begin{center}
\includegraphics[scale=0.6] {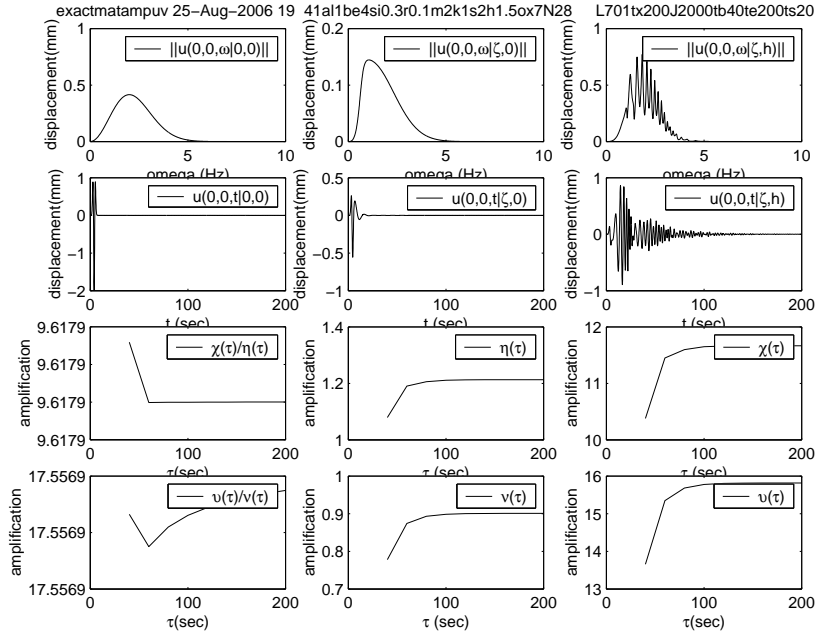}
\end{center}
\caption{$h=1.5$ . Large-scale anomalous effects} \label{graf2}
\end{figure}

The results of the computations, peformed by direct resolution of
a suitably-chosen finite dimension version of the matrix equation,
are given in figs. \ref{graf1} and \ref{graf2} for $h=0.1$ and
$h=1$ respectively. The first line in these figures contains the
spectra, the second line the time histories, and the remaining
lines the ratios of cumulative response.
%
\section{Comments}
The spectra reveal the existence of a {series of resonances}, as
predicted by the theory.

The peak displacement (in the time domain) on a
spatially-modulated impedance ground is always smaller than the
peak displacement on the flat, stress-free ground.

The maximum values of the ratios $\eta$, $\chi$, $\nu$ and
$\upsilon$ first increase and then decrease with increasing $h$,
with the {maximum amplifications ($\eta\approx 1.2$, $\chi\approx
11.7$, $\nu\approx 0.9$, $\upsilon\approx 15.7$)} being attained
for $h=1.5$.

It is possible to obtain substantial amplification of the ratios
with respect to the spatially-constant impedance flat surface and
much less amplification or even deamplification of the ratios with
respect to the stress-free flat surface.

This shows that although there is no doubt that
spatially-modulated impedance surfaces can give rise to {beating
and very long durations, attaining in some of these examples of
the order of 2 min} versus of the order of 10-20 sec for zero or
non-zero constant impedances), there is {no clear-cut answer to
the question of whether spatial impedance modulations, which
simulate the existence of ground uneveness, systematically result
in amplification or even deamplification of cumulative ground
motion displacement and velocity, especially with respect to
motion of a stress-free flat ground}.
\bibliographystyle{plain}
\bibliography{biblio_seismes_2}
\end{document}